\documentclass[mathleft]{an}
\usepackage{times}
\usepackage{graphicx}
\usepackage{url}
\usepackage{bm}
\graphicspath{{./fig/}{./png/}}

\newcommand{\EQ}{\begin{equation}}
\newcommand{\EN}{\end{equation}}
\newcommand{\EQA}{\begin{eqnarray}}
\newcommand{\ENA}{\end{eqnarray}}

\newcommand{\Eq}[1]{Eq.~(\ref{#1})}

\newcommand{\Sec}[1]{Sect.~\ref{#1}}

\newcommand{\Fig}[1]{Fig.~\ref{#1}}
\newcommand{\FFig}[1]{Figure~\ref{#1}}

\newcommand{\bra}[1]{\langle #1\rangle}

\newcommand{\meanFF}{\overline{\mbox{\boldmath ${\cal F}$}} {}}
\newcommand{\meanA}{\overline{A}}
\newcommand{\meanB}{\overline{B}}
\newcommand{\meanJ}{\overline{J}}
\newcommand{\meanU}{\overline{U}}

\newcommand{\meanAA}{\overline{\vec{A}}}
\newcommand{\meanBB}{\overline{\vec{B}}}
\newcommand{\meanJJ}{\overline{\vec{J}}}
\newcommand{\meanUU}{\overline{\vec{U}}}
\newcommand{\meanWW}{\overline{\vec{W}}}

\newcommand{\meanSSSS}{\overline{\mbox{\boldmath ${\mathsf S}$}} {}}
\newcommand{\meanSSS}{\overline{\mathsf{S}}}
\newcommand{\ggamma}{{\vec{\gamma}}}

\newcommand{\uu}{{\vec{u}}}

\newcommand{\nab}{\mbox{\boldmath $\nabla$} {}}

\newcommand{\DD}{{\rm D} {}}

\def\Rm{R_{\rm m}}
\def\Pm{P_{\rm m}}
\def\Rey{\mbox{\rm Re}}

\def\Sh{\mbox{\rm Sh}}

\def\kf{k_{\rm f}}
\def\etat{\eta_{\rm t}}

\def\half{{\textstyle{1\over2}}}

\def\onethird{{\textstyle{1\over3}}}

\newcommand{\km}{\,{\rm km}}

\newcommand{\Mx}{\,{\rm Mx}}

\newcommand{\yapj}[3]{: #1, {ApJ} {#2}, #3}

\newcommand{\yan}[3]{: #1, {AN} {#2}, #3}

\newcommand{\yana}[3]{: #1, {A\&A} {#2}, #3}

\newcommand{\ygafd}[3]{: #1, {GApFD} {#2}, #3}

\newcommand{\yprl}[3]{: #1, {Phys Rev Lett} {#2}, #3}
\newcommand{\ypre}[3]{: #1, {Phys Rev E} {#2}, #3}

\newcommand{\ymn}[3]{: #1, {MNRAS} {#2}, #3}
\newcommand{\ynat}[3]{: #1, {Nature} {#2}, #3}

\newcommand{\yjour}[4]{: #1, {#2} {#3}, #4}

\Pagespan{725}{}
\Yearpublication{2008}
\Yearsubmission{2008}
\Month{6}
\Volume{329}
\Issue{7}
\DOI{10.1002/asna.200811027}

\begin{document}

\title{The dual role of shear in large-scale dynamos}
\authorrunning{A. Brandenburg}
\author{A. Brandenburg\thanks{Corresponding author: brandenb@nordita.org   }}
\institute{
NORDITA, Roslagstullsbacken 23, SE-10691 Stockholm, Sweden
}

\received{2008 Jun 2}  \accepted{2008 Jul 18}
\publonline{2008 Aug 30}


\abstract{%
The role of shear in alleviating catastrophic quenching by shedding
small-scale magnetic helicity through fluxes along contours of constant
shear is discussed.
The level of quenching of the dynamo effect depends on the quenched
value of the turbulent magnetic diffusivity.
Earlier estimates that might have suffered from the force-free
degeneracy of Beltrami fields are now confirmed for shear flows where
this degeneracy is lifted.
For a dynamo that is saturated near equipartition field strength
those estimates result in a 5-fold decrease of the magnetic diffusivity
as the magnetic Reynolds number based on the wavenumber of the
energy-carrying eddies is increased from 2 to 600.
Finally, the role of shear in driving turbulence and large-scale fields
by the magneto-rotational instability is emphasized.
New simulations are presented and the $3\pi/4$ phase shift between
poloidal and toroidal fields is confirmed.
It is suggested that this phase shift might be a useful diagnostic tool in
identifying mean-field dynamo action in simulations and to distinguish
this from other scenarios invoking magnetic buoyancy as a means to
explain migration away from the midplane.
}

\maketitle

\section{Introduction}

Shear clearly plays an important role in amplifying toroidal fields
from poloidal, but that is not all.
Shear also plays a role in ``unquenching'' any dynamo effect that may
play a role in producing poloidal field from toroidal, thus closing
the dynamo loop.
A prime example of such a dynamo effect is the $\alpha$ effect, but
other possible known effects may include the shear-current effect
and the incoherent alpha-shear effect.
The ``unquenching'' of dynamo effects, as well as the dynamo effects
themselves, require more detailed considerations of the results
available so far.
This is the principal goal of this paper.

The term ``unquenching'' in connection with the $\alpha$ effect
may appear somewhat unusual, but this choice of words must be seen
in contrast to the possibility of catastrophic $\alpha$ quenching.
Here, ``catastrophic'' indicates that the quenching by the mean
magnetic field, $\meanBB$, becomes more extreme as the magnetic
Reynolds number, $\Rm$, increases.
Traditionally, such quenching is represented by the simplistic formula
(Vainshtein \& Cattaneo 1992)
\EQ
\alpha(\meanBB)={\alpha_0\over1+\Rm\meanBB^2/B_{\rm eq}^2},
\label{QuenchSimple}
\EN
where $B_{\rm eq}=\bra{\mu_0\rho\uu^2}^{1/2}$ is the equipartition
field strength with respect to the kinetic energy density, $\uu$ is
the small-scale turbulent velocity, $\rho$ is the density, $\mu_0$
is the vacuum permeability, $\Rm=u_{\rm rms}/\eta\kf$ is the magnetic
Reynolds number, $\eta$ is the magnetic diffusivity,
$\kf$ is the wavenumber of the energy-carrying scale,
and angular brackets denote volume averaging.
However, \Eq{QuenchSimple} is really only valid under special
circumstances that are quite uninteresting for dynamo action:
infinite wavelength of the magnetic field, complete stationarity,
and no possibility of magnetic helicity fluxes.
The latter possibility is now believed to be the most important one for
astrophysical dynamos, as was first suggested by Blackman \& Field (2000).
Any one of these three caveats alleviates the severity of catastrophic
quenching, because they all lead to ``extra terms'' that enter in the
numerator of \Eq{QuenchSimple} with an $\Rm$ factor in front, i.e.\
\EQ
\alpha(\meanBB)={\alpha_0+\Rm\times\mbox{``extra effects''}
\over1+\Rm\meanBB^2/B_{\rm eq}^2},
\label{QuenchExtra}
\EN
where we have assumed that the kinematic $\alpha$ value, $\alpha_0$,
remains independent of time.
In its essence, this equation with extra effects included goes back to the early work of
Kleeorin \& Ruzmaikin (1982), and later Kleeorin et al.\ (1995, 2000),
and has been discussed in connection with alleviating catastrophic
$\alpha$ quenching by Blackman \& Brandenburg (2002)
and Brandenburg \& Subramanian (2005a).
A more complete quenching formula with extra effects included takes
the form
\EQ
\alpha={\alpha_0+\Rm\left(
\eta_{\rm t}{\mu_0\meanJJ\cdot\meanBB\over B_{\rm eq}^2}
-{\nab\cdot\meanFF_{\rm C}\over2 k_{\rm f}^2B_{\rm eq}^2}
-{\partial\alpha/\partial t\over2\eta_{\rm t} k_{\rm f}^2}\right)
\over1+\Rm\meanBB^2/B_{\rm eq}^2}.
\label{QuenchExtra2}
\EN
Let us now discuss separately all three terms in the parenthesis
of the numerator of \Eq{QuenchExtra2}.

(i) The effect of the $\meanJJ\cdot\meanBB$ term is clearly seen when
considering the saturation of homogeneous dynamos in a periodic domain.
In that case this formula gives
\EQ
\alpha(\meanBB)\rightarrow\eta_{\rm t}\mu_0\meanJJ\cdot\meanBB/\meanBB^2
\quad\mbox{(for $\Rm\to\infty$)},
\EN
so there is nothing catastrophic about this formula,
unless $\eta_{\rm t}$ itself is catastrophically quenched.
Of course, if mean fields are defined as full volume averages,
$\meanBB$ becomes completely uniform, so $\mu_0\meanJJ=\nab\times\meanBB=0$.
This is the case in numerical experiments by Cattaneo \& Hughes (1996).
The other case with a finite $\meanJJ\cdot\meanBB$ term was seen in the
simulations of Brandenburg (2001), where not only a large-scale magnetic
field was found to saturate at super-equipartition values, but also the
$\alpha$ and $\eta_{\rm t}$ effects were found to be only mildly quenched.

(ii) The $\partial/\partial t$ term in \Eq{QuenchExtra2} is important to
explain the absence of an otherwise premature onset of quenching at
resistively low field strengths where $\meanBB^2/B_{\rm eq}^2\approx\Rm^{-1}$.
This is also confirmed by controlled numerical experiments where the initial
field was a weak Beltrami field (Brandenburg et al.\ 2003).

\begin{figure}[t!]\begin{center}
\includegraphics[width=\columnwidth]{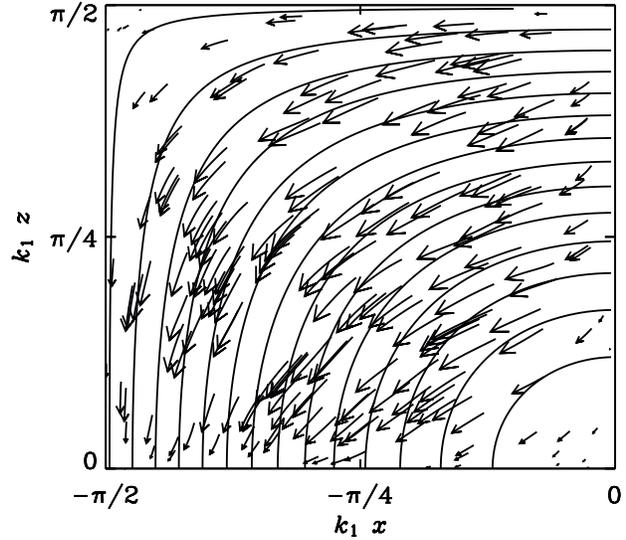}
\end{center}\caption[]{ 
Vectors of $\meanFF_{\rm C}$ together with contours of $\meanU_y$ which
also coincide with the streamlines of the mean vorticity field $\meanWW$.
Note the close agreement between $\meanFF_{\rm C}$ vectors and
$\meanWW$ contours.
The orientation of the vectors indicates that negative current helicity
leaves the system at the outer surface ($x=0$).
Adapted from Brandenburg et al.\ (2005).
}\label{p2vishcho}\end{figure}

(iii) Finally, the effect of magnetic helicity fluxes was first seen in
simulations with imposed fields by Brandenburg \& Sandin (2004), and
then later in dynamo simulations (Brandenburg 2005a), which brings us to the
main topic of this paper.
As was already seen in earlier simulations without shear, just allowing
for open boundary conditions alone does not help to produce a finite
magnetic helicity flux in \Eq{QuenchExtra2} and hence does not alleviate
catastrophic $\alpha$ quenching (Brandenburg \& Dobler 2001).
However, Vishniac \& Cho (2001) showed that in the presence of
differential rotation a magnetic helicity flux can be generated and that
it flows along the rotation axis.
More detailed work of Subramanian \& Brandenburg (2004, 2006) and
Brandenburg \& Subramanian (2005b) resulted in a simple formula for
this particular contribution to the flux:
\EQ
\meanFF_{\rm C}=C_{\rm VC}(\meanSSSS\,\meanBB)\times\meanBB,
\EN
where $\meanSSS_{ij}=\half(\meanU_{i,j}+\meanU_{j,i})$
is the rate of strain matrix of the mean flow, and $C_{\rm VC}$ is a
dimensionless number of order unity.
This flux is along contours of constant shear, as was demonstrated
by Brandenburg et al.\ (2005); see also \Fig{p2vishcho}.

The sign of the magnetic helicity flux is negative in the Northern
hemisphere, so vectors of positive magnetic helicity flux point away
from the surface in the Northern hemisphere, then through the equator
and into the Southern hemisphere.

\section{Quantitative considerations}
\label{Incoherent}

Simulations of Brandenburg \& Sandin (2004) indicate that the 
magnitude of the Vishniac-Cho flux might be on the order of
$10^{46}\Mx^{24}$/cycle, if applied to the Sun.
In units of $\meanFF_0\equiv u_{\rm rms}k_{\rm f}B_0^2$ the
nondimensional flux was estimated to be about $30/\half\Sh$,
where
\EQ
\Sh\equiv S/u_{\rm rms}k_{\rm f}
\EN
is the nondimensional shear parameter and $S$ is the shear rate.
The corresponding $C_{\rm VC}$ parameter would then be $\gg1$.
This result appears in conflict with the expectation that $C_{\rm VC}$
should be of order unity (Subramanian \& Brandenburg 2004;
Brandenburg \& Subramanian 2005b).
It would therefore be important to return to this issue using
a simpler one-dimensional shear profile together with open
boundary conditions.

\begin{figure}[t!]\begin{center}
\includegraphics[width=\columnwidth]{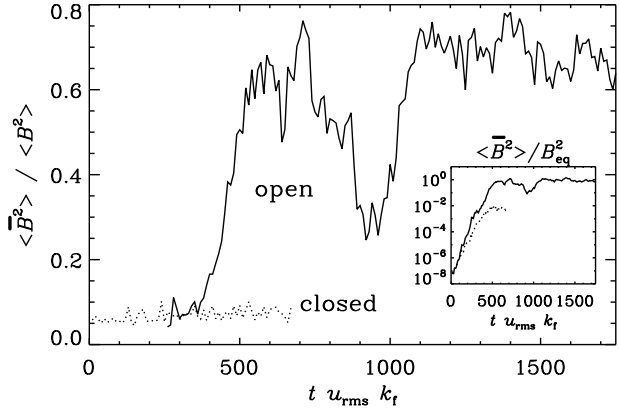}
\end{center}\caption[]{ 
Evolution of the ratio of the energies contained in the large-scale field
to the total magnetic energy for open and closed boundaries.
Note that large-scale dynamo action is only possible with open
(vertical field) boundary conditions.
Adapted from Brandenburg (2005a).
}\label{pmean_comp_catania}\end{figure}

A more conclusive indication for the operation of he\-li\-ci\-ty fluxes is the
demonstration of a successful dynamo simulation; see, e.g.,
Brandenburg (2005a), where it was also demonstrated that with closed
perfect conductor boundary conditions the dynamo was not successful
in producing large-scale fields (at least not during the course of the
present simulation).
In \Fig{pmean_comp_catania} we plot the ratio of
the energy contained in the large-scale field to the total
magnetic energy.
We see clearly that large-scale dynamo action is only possible
with open boundary conditions.
By open we mean here the vertical field condition that is commonly used
in simulations of magneto-convection (Hurlburt \& Toomre 1988).
Such boundary conditions do permit a finite magnetic helicity flux, but they
do not allow Poynting flux to pass through the boundaries.
It is at present unclear whether the absence of a Poynting flux
is a serious short-coming or not.

Another remarkable result found in the work of Brandenburg (2005a) is
that large-scale magnetic fields can be found even without kinetic helicity
and hence without $\alpha$ effect.
An obvious possibility might be that this is caused by the shear-current
effect of Rogachevskii \& Kleeorin (2003, 2004).
However, in the case of a simpler geometry where the contours of constant
shear are purely vertical, which is relevant for accretion discs, for example,
no direct support for the existence of this effect has been found
(Brandenburg 2005b; Brandenburg et al.\ 2008a).
This is also consistent with earlier analytic results of
R\"udiger \& Kitchatinov (2006) and R\"adler \& Stepanov (2006) using
the second order correlation approximation (SOCA).
In essence, a shear-current effect would be described by a mean-field
equation with an anisotropic magnetic diffusion tensor, $\eta_{ij}$.
We write the governing equation here for the magnetic vector potential,
\EQ
{\DD\meanA_i\over\DD t}=-\meanA_j\meanU_{j,i}-\mu_0\eta_{ij}\meanJ_j-\mu_0\eta\meanJ_i,
\EN
where $\meanJJ=-\nabla^2\meanAA$, and
one-dimensional averages have been employed, i.e.\
$\meanJJ=\meanJJ(z,t)$ in the present case, and $\meanUU=(0,Sx,0)$.
In the case of vertical contours of the mean shear, we have $\meanU_{2,1}=S$
and all other components vanish.
Here, the cross-stream direction is $i=x$ or 1, and the streamwise
direction is $i=y$ or 2.

\begin{figure}[t!]\begin{center}
\includegraphics[width=\columnwidth]{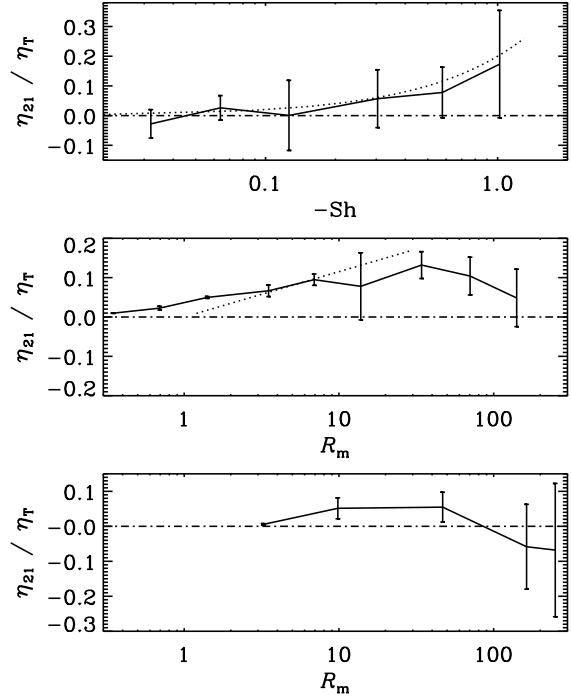}
\end{center}\caption[]{ 
Dependence of $\eta_{21}$, normalized by the total (turbulent and
microscopic) magnetic diffusivity on shear parameter (upper panel)
and on magnetic Reynolds number for $\Rey=1.4$ (middle panel)
and for a fixed magnetic Prandtl number, $\Pm=20$ (lower panel).
Adapted from Brandenburg et al.\ (2008a).
}\label{kinshear_summary}\end{figure}

In order to have a closed dynamo loop, one would then need to have a finite
$\eta_{21}$ component with the same sign as that of $S$.
According to SOCA calculations and simulations this is however not the case.
In \Fig{kinshear_summary} we show the resulting values of $\eta_{21}$
for such a linear shear flow, as obtained using the testfield method.

Here we have used $S<0$, so we are looking for negative values of
$\eta_{21}$ for an operational shear-current effect.
In all cases we find $\eta_{21}>0$, except for large values of
$\Rm$ when $\Pm=20$.
However, the error bars are large.
There is perhaps the possibility that the sign of $\eta_{21}$ may change
under other circumstances, e.g.\ for the more complicated shear profile
shown in \Fig{p2vishcho}.
Yet another possibility is that in the presence of helicity the sign
may change.
Some evidence to this effect has been provided by Mitra et al.\ (2008).
It turns out that in nonhelical shear flow turbulence, $\eta_{21}$
can become negative in the saturated state with helicity.
However, given that there is still an $\alpha$ effect, the relevance
of the shear-current effect is less obvious in these simulations.

Obviously, with helicity there is also an $\alpha$ effect, so the
shear-current effect would not be the sole cause of large-scale
dynamo action.
This leaves us with the question what causes large-scale magnetic field
generation in non-helical turbulence with shear?
In addition to the simulations of Brandenburg (2005a), discussed above,
there are also simulations of Yousef et al.\ (2008), where a large-scale
magnetic field is generated.
They find a growth rate that is proportional to the shear rate $S$.
The authors argue that such a result would not be consistent with the
shear-current effect, because the growth rate would then be proportional
to the product of $S$ and $\eta_{21}$, where $\eta_{21}$ itself would
be proportional to $S$.

\begin{figure}[t!]
\centering\includegraphics[width=\columnwidth]{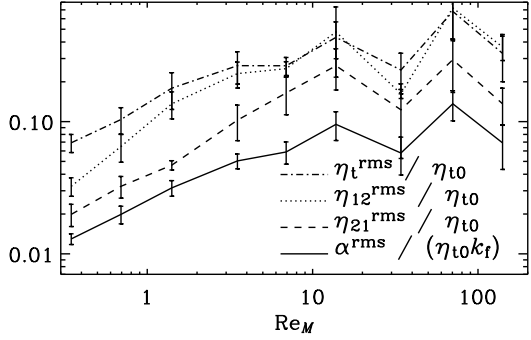}\caption{
Dependences of the rms values of the temporal fluctuations
$\alpha_{\rm rms}$ (normalized by $\eta_{\rm t0}k_{\rm f}$),
$\eta_{\rm t}^{\rm rms}$, $\eta_{21}^{\rm rms}$, and $\eta_{12}^{\rm rms}$
(normalized by $\eta_{\rm t0}$), on $\Rm$
for $\Rey=1.4$ and $\mbox{Sh}=-0.6$.
Adapted from Brandenburg et al.\ (2008a).
}\label{kinshear_Re14_alprms}\end{figure}

The alternative proposal by Brandenburg et al.\ (2008a) is that an
incoherent $\alpha$-shear or $\alpha\Omega$ dynamo is at work.
The occurrence of an incoherent $\alpha$-shear dynamo
(cf.\ Vishniac \& Brandenburg 1997; Proctor 2007; Kleeorin \& Rogachevskii 2008)
was quantified by estimating the rms values of the fluctuations
of all components of $\alpha_{ij}$ and $\eta_{ij}$.
\FFig{kinshear_Re14_alprms} shows how these rms values vary with
increasing values of $\Rm$.
We recall that the average of all components
of $\alpha_{ij}$ is zero, so there is no regular $\alpha$ effect.
The onset of incoherent dynamo action depends on the value of the dynamo number
\EQ
D_{\alpha S}^{\rm incoh}=\alpha_{\rm rms} S/\eta_{\rm T}k_1^3,
\EN
where $\alpha_{\rm rms}$ is the rms value of the streamwise
component $\alpha_{22}$
(but all components are found to have the same rms value),
$\eta_{\rm T}=\eta_{\rm T}+\eta$ is the sum of turbulent and microscopic
magnetic diffusivity, and $k_1=2\pi/L$ is the lowest wavenumber in the
$z$ direction of the box.

The simulations of Brandenburg et al.\ (2008a) give fluctuations of
$\alpha_{\rm rms}$ that correspond to $D_{\alpha S}^{\rm incoh}\approx4$.
This is to be compared with the marginal value of $\approx2.3$ obtained
numerically using a single-mode approximation.
There is in principle also the incoherent shear-current effect, based
on the random occurrence of values of $\eta_{21}$ with suitable sign.
However, the effect is found to be subdominant compared with the dynamo
number of the incoherent $\alpha$-shear dynamo.
Yet another possibility for generating large-scale magnetic fields in the
presence of shear is given by additive noise in the electromotive force
due to small-scale dynamo action (Blackman 1998).
However, in Yousef et al.\ (2008) no small-scale dynamo was excited, so
this type of noise would not explain the field seen in their simulation.

In all cases one would expect random reversals of the toroidal magnetic
field on a turbulent diffusion time scale.
However, simulations usually give longer time scales, which may be
explained by a tendency to conserve magnetic helicity.
Indeed, using  magnetic helicity conservation, Brandenburg et al.\
(2008a) found that in a closed domain the reversal time increases with
$\Rm$ to the 1/2 power.
The mean reversal time might therefore well be much larger than
a turbulent diffusion time.

\section{Shear in helical turbulent dynamos}

\begin{figure}[t!]\begin{center}
\includegraphics[width=\columnwidth]{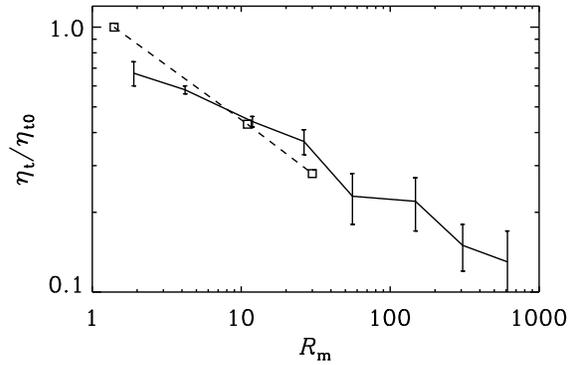}
\end{center}\caption[]{ 
$R_{\rm m}$ dependence of $\etat$ in the dynamo-saturated state
as measured by the testfield method (solid line; Brandenburg et al.\ 2008b),
and estimated for oscillatory $\alpha\Omega$ dynamos in a shearing box
(dashed line; K\"apyl\"a \& Brandenburg, in preparation).
}\label{peta}\end{figure}

Let us now look at the effects of shear in simulations where kinetic helicity
and thus an $\alpha$ effect are present.
In these cases there is the possibility of oscillatory solutions
with frequency $\omega_{\rm cyc}$ and propagating dynamo waves
of wavenumber $k$ and  wave speed $c=\omega_{\rm cyc}/k$.
The sense of propagation is determined by the sign of the product
$\alpha S$.
When shear is sufficiently strong ($|S/\alpha k|\gg1$ or,
equivalently, $|\Sh|\gg\onethird k_1/\kf$ for fully helical turbulence
with $\alpha\approx\onethird u_{\rm rms}$), the magnitude of the
oscillation frequency is given by $(\alpha k S/2)^{1/2}$, but in the
marginal state this must be balanced by the diffusion rate $\eta_{\rm T} k^2$.
As was stressed by Blackman \& Brandenburg (2002), this provides therefore
a robust tool for determining empirically the quenched value of
$\eta_{\rm T}=\eta_{\rm t}+\eta$.
However, it has been rather hard to reach large enough values of the
ratio $\eta_{\rm t}/\eta$ (which is also a proxy of $\Rm$) to make conclusive
statements about the $\Rm$-dependence of $\eta_{\rm t}$.
Indeed, the values obtained so far (\Fig{peta}) confirm recent results of
Brandenburg et al.\ (2008b), where a measure of $\eta_{\rm  t}$ has been
obtained using the testfield method applied to the nonlinear state
for $\Rm$ ranging between 2 and 600.
Unfortunately, those results leave some ambiguity between the
$\alpha$ effect and the turbulent magnetic diffusivity, because
of terms of the form $(\meanJJ\cdot\meanBB)\meanBB$, whose coefficients
can be associated both with $\alpha$ and with $\eta_{\rm t}$.

\section{MRI-driven turbulence}

\begin{figure}[t!]\begin{center}
\includegraphics[width=\columnwidth]{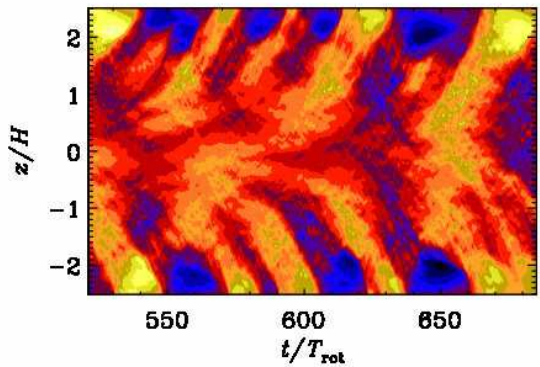}
\end{center}\caption[]{ 
(online colour at: www.an-journal.org) Space-time ($z$--$t$) diagram of $\meanB_y(z,t$ for a run where
turbulence is driven by the magneto-rotational instability.
The magnetic Reynolds number, based on the actual rms velocity and the
pseudo wavenumber $k_{\rm f}=2\pi/H$, is about 120.
}\label{64x256c8_test}\end{figure}

Interesting systems where the effects of shear play an absolutely
vital role are seen are local simulations of
accretion disc turbulence that is driven by the magneto-rotational
instability (Hawley et al.\ 1995).
When there is also stratification about the midplane, the Coriolis force
produces an $\alpha$ effect.
Simulations of Brandenburg et al.\ (1995) give rise to
large-scale dynamo action that turns out to be oscillatory with dynamo waves
propagating away from the midplane.
In \Fig{64x256c8_test} we show new simulations that are similar to those of
Brandenburg et al.\ (1995), but with somewhat larger resolution and without
using hyperviscosity.
Only shock viscosity and shock resistivity are being used.
Another difference is that these simulations also have a potential field
boundary condition instead of the vertical field condition that was
used in Brandenburg et al.\ (1995).
Given the remarkable similarity with the earlier simulations, we must
conclude that the different boundary conditions do not seem to have a
major effect.
The new simulations are similar in that they too show oscillations with
a typical period of about 30--50 orbits (the orbital time is defined
as $T_{\rm rot}=2\pi/\Omega$).
However, the new simulations also show considerably more fluctuations
with parity variations and, more importantly, a noticeable decoupling
of behavior in the Northern and Southern disc planes.
The parity varies therefore between more nearly symmetric (even) and
more nearly antisymmetric (odd) parity.

\begin{figure}[t!]\begin{center}
\includegraphics[width=\columnwidth]{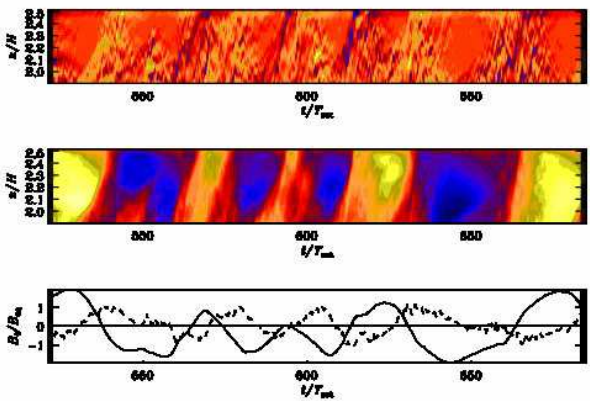}
\end{center}\caption[]{ 
(online colour at: www.an-journal.org) Mean poloidal and toroidal fields in a narrow slice near the upper
disc plane as a function of $t$ and $z$, together with the averages
over the same slice showing the phase relation between the two.
The solid line indicates $\meanB_y/B_{\rm eq}$ while the dotted
line indicates $\meanB_x/B_{\rm eq}$, scaled up by a factor of 20
to make it better visible.
}\label{phase_64x256c8_test}\end{figure}

\begin{figure}[t!]\begin{center}
\includegraphics[width=\columnwidth]{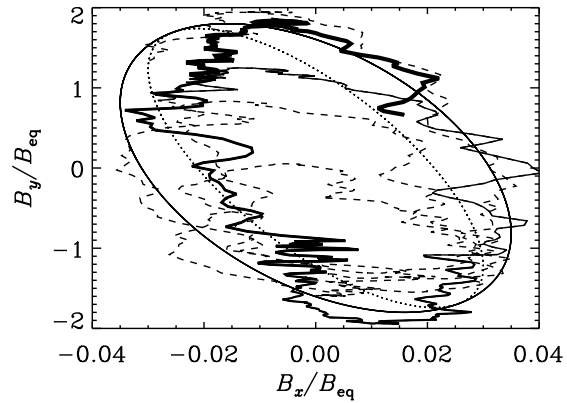}
\end{center}\caption[]{ 
Phase plot of the averages of poloidal and toroidal fields
over the same narrow slices as in \Fig{phase_64x256c8_test}.
The first times are plotted as dashed lines, but the later
times are solid with increasing thickness toward the end
showing that a point on the curve moves forward in a clockwise
direction.
Overplotted are two ellipses showing $B_{y0}\propto\cos(\omega t+\phi)$ versus
$B_{x0}\propto\cos\omega t$ with $\phi=0.65\pi$, $B_{x0}=0.035\,B_{\rm eq}$,
and $B_{x0}=1.8\,B_{\rm eq}$ (solid line) and $\phi=0.75\pi$,
$B_{x0}=0.03\,B_{\rm eq}$, and $B_{x0}=1.74\,B_{\rm eq}$ (dotted line).
}\label{phase2_64x256c8_test}\end{figure}

\section{Detecting mean-field dynamo action}

A possible means of identifying $\alpha\Omega$-type dynamo action
as the main course of oscillations seen in simulations we propose
to determine the phase relation between poloidal and toroidal
fields.
This used to be a standard tool in solar dynamo theory to infer the
sense of radial differential rotation, but may also become an
important tool in disc and other oscillatory dynamos.
Mean-field theory predicts a phase shift by ${3\over4}\pi$,
which was confirmed by Brandenburg \& Sokoloff (2002).
Another alternative explanation for the migration away from the
midplane would be magnetic buoyancy.
This was discussed by Vishniac \& Brandenburg (1997), who
noted that the migration speed is only about 3\% of the
turbulent rms velocity.
The idea of magnetic buoyancy playing a leading role was expressed
again in connection with recent simulations of Blaes et al.\ (2008),
where the outer disc surface is marked by a surface where the optical
depth in a radiative transfer calculation was of order unity.
However, no detailed proposal for the phase relation from the
buoyancy effect has yet been made.

\begin{figure}[t!]\begin{center}
\includegraphics[width=\columnwidth]{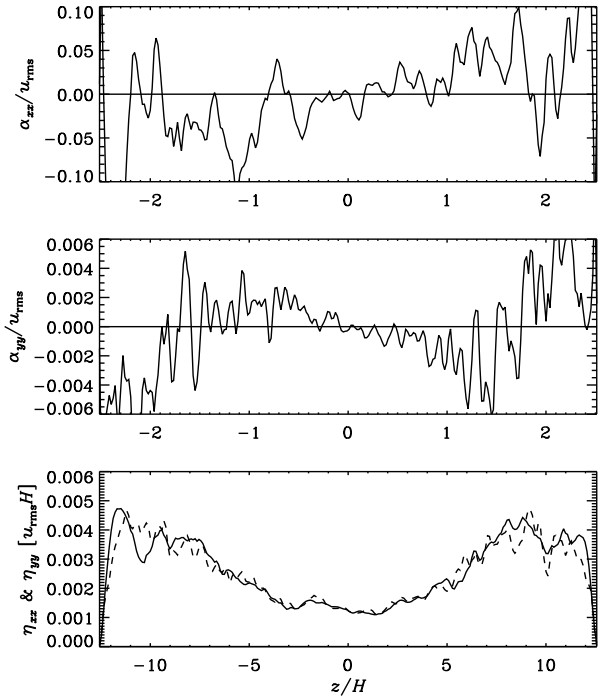}
\end{center}\caption[]{ 
Vertical profiles of $\alpha_{xx}$, $\alpha_{yy}$ (first and
second panel), both normalized by the turbulent rms velocity,
as well as $\eta_{xx}$ (solid line) and $\eta_{yy}$ (dashed
line) for an MRI disc simulations.
}\label{64x256c8_test_alpetaijm}\end{figure}

Brandenburg et al.\ (1995) found that the details of the large-scale
magnetic field generation can be described by an $\alpha$ effect that
is negative in the Northern disc plane.
Newer determinations of $\alpha_{ij}$ and $\eta_{ij}$ using the testfield
method confirm this result also for the present simulations;
see \Fig{64x256c8_test_alpetaijm}.
Again, we have here access to all 4+4 components of the $\alpha_{ij}$
and $\eta_{ij}$ tensors.
As in Brandenburg \& Sokoloff (2002), who used a correlation method
instead of the testfield method, $\alpha_{yy}$ is negative in the
upper midplane and has an extremum at $z\approx\pm H$.
However, $\alpha_{11}$ ($\equiv\alpha_{xx}$) is positive in the upper
disc plane.

\begin{figure}[t!]\begin{center}
\includegraphics[width=\columnwidth]{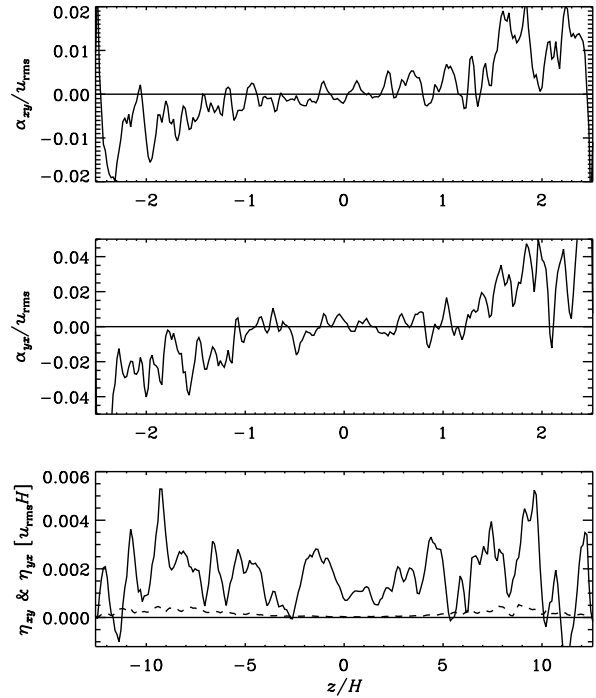}
\end{center}\caption[]{ 
Vertical profiles of $\alpha_{xy}$, $\alpha_{yx}$ (first and
second panel), both normalized by the turbulent rms velocity,
as well as $\eta_{xy}$ (solid line) and $\eta_{yx}$ (dashed
line) for an MRI disc simulations.
}\label{64x256c8_test_alpetaijm2}\end{figure}

In all cases we find that $\eta_{11}\approx\eta_{22}$ and always positive,
in contrast to earlier work using the correlation method.
This function has a minimum in the midplane and grows away from
the midplane by a factor of about 4.
The off-diagonal components of $\alpha_{ij}$ and $\eta_{ij}$ are
shown in \Fig{64x256c8_test_alpetaijm2}.

It turns out that, like $\alpha_{xx}$ and $\alpha_{yy}$,
also $\alpha_{xy}$ and $\alpha_{yx}$ are approximately antisymmetric
about the midplane and positive in the upper disc disc plane.
The off-diagonal components of $\alpha_{ij}$ are normally interpreted
in terms of turbulent pumping with an effective vertical velocity
$\ggamma=\half(\alpha_{yx}-\alpha_{xy})$.
However, since $\alpha_{yx}$ and $\alpha_{xy}$ have the same sign,
there is some cancelation.
Nevertheless, since $\alpha_{yx}$ is larger than $\alpha_{xy}$, there
is net transport away from the midplane.
This concerns predominantly the toroidal field component, $\meanB_y$,
while the poloidal component, $\meanB_x$, is transported predominantly
toward the midplane.
Such differential pumping of poloidal and toroidal fields was first
discussed by Kitchatinov (1991) and later confirmed in simulations by
Ossendrijver et al.\ (2002).

The off-diagonal components of $\eta_{ij}$ are symmetric about the midplane
and both are mainly positive.
The $\eta_{yx}$ component is important for the shear-current effect,
but it is found to have the wrong sign for this effect
to operate in our simulations; see also \Sec{Incoherent}.

\section{Dynamos in shearing convection}

Large-scale dynamo action of significant amplitude has been seen in
global simulations of the geodynamo (Glatzmaier \& Roberts 1995).
However, simulations in Cartesian boxes have not yet shown such behavior.
A recent example is that by Tobias et al.\ (2008) where strong shear was
present.  However, the absence of large-scale fields of significant magnitude
may be related to the orientation of shear.
In their case those contours are horizontal, so the Vishniac-Cho flux,
that goes along those contours, would not be able to escape the
horizontally periodic domain.
If it is true that large-scale dynamo action only works efficiently
if excess small-scale magnetic helicity is expelled through the boundaries,
then it would be plausible that shear may not help the dynamo in all cases.
More recent convection simulations by K\"apyl\"a et al.\ (2008)
do show large-scale magnetic field in the presence of shear.
They have contours of constant shear that do cross the boundaries.
In fact, their shear is just like that in the accretion disc simulations
discussed earlier.

\section{Discussion}

In this paper we have illustrated the important dual role played by
shear: helping to produce strong toroidal field and helping to
unquench the $\alpha$ effect.
We have tried to make the case that successful large-scale dynamos
must have an opportunity to shed small-scale magnetic helicity through
the boundaries.
This is what the Sun does (e.g., D\'emoulin et al.\ 2002),
and this is also what many simulations can do,
although the degree of realism to which simulations can do this varies.
So far, there is no simulation that goes all the way to including at least
a simplified way of modeling coronal mass ejections as the final stage
in shedding small-scale magnetic helicity.
This should obviously be tried in the near future using simulations
in spherical shell segments, possibly with an additional outer layer
where wind acceleration can occur (Brandenburg et al.\ 2007).

In this connection it might be helpful to explain the special meaning
of ``small scale'' in connection with large-scale dynamos.
In solar physics, one is used to associating active regions with large
scales, while small scales would often refer to the resolution limit
of $100\km$ or less.
Obviously, in connection with the solar cycle, relevant time and length
scales would be on the order of years and several hundred megameters,
respectively -- well encompassing the duration and size of active regions.
In the present work we have used averages that do not a priori separate
between large and small scales.
By assuming an average over one or two coordinate directions (e.g.\ an
azimuthal average for global simulations of the Sun, or horizontal
averages in local box simulations), there could still be residual
fluctuations on short time and small length scales, but they should
go away with increasing system size in the azimuthal or horizontal
directions like one over the square root of the number of eddies that
are being averaged over.
Exactly this fact is behind the operation of the incoherent $\alpha$-shear
dynamo discussed at the end of \Sec{Incoherent}.
Clearly a finite horizontal extent is not only a numerical restriction of
local boxes.
Indeed, a finite azimuthal extent is bare reality even for the Sun.
This point has been made by Hoyng (1993) in order to explain limits to
the phase and amplitude stability of the solar cycle, but this very
mechanism could even suffice to drive a dynamo in shearing systems
where there is no $\alpha$ effect present.
It should be emphasized that the incoherent $\alpha$ effect as such
is independent of shear, but it is able to produce coherent large-scale
fields only in the presence of shear.
Obviously, more work is needed to sharpen the analytical and numerical
tools to understand such systems better.
Finally, let us emphasize that incoherent $\alpha$-effect dynamos too
have a nonlinear stage
that is controlled by magnetic helicity at some level.
In Brandenburg et al.\ (2008a) we only used a very simplistic one-mode
truncation to make this point, but the whole dynamical quenching
formalism applies otherwise just as well.

\acknowledgements
I thank Eric G. Blackman for comments on this paper and
the organizers of the KITP program on dynamo theory for providing
a stimulating atmosphere during its programs on dynamo theory.
This research was supported in part by the National Science
Foundation under grant PHY05-51164.
Computational resources were made available by the 
Swedish National Allocations Committee at the
National Supercomputer Centre in Link\"oping.

\end{document}